\documentclass[aps,pre,onecolumn,superscriptaddress,showpacs,showkeys]{revtex4}
\usepackage[]{graphicx}
\begin{document}

\title{Thermal noise suppression: How much does it cost?}
\author{Giovanni Volpe}
\affiliation{ICFO - Institut de Ciencies Fotoniques,
Mediterranean Technology Park, 08860, Castelldefels (Barcelona),
Spain}
\author{Jan Wehr}
\affiliation{ICFO - Institut de Ciencies Fotoniques,
Mediterranean Technology Park, 08860, Castelldefels (Barcelona),
Spain}
\affiliation{Department of Mathematics, University of Arizona, Tucson, Arizona 85721-0089, USA}
\author{Dmitri Petrov}
\affiliation{ICFO - Institut de
Ciencies Fotoniques, Mediterranean Technology Park, 08860,
Castelldefels (Barcelona), Spain} \affiliation{ICREA -
Instituci\'{o} Catalana de Recerca i Estudis Avan\c{c}ats, 08010,
Barcelona, Spain}
\author{J. Miguel Rubi}
\affiliation{Departament de F\'{i}sica Fonamental, Universitat de Barcelona, Diagonal 647,
08028 Barcelona, Spain}
\date{\today}

\begin{abstract}
In order to stabilize the behavior of noisy systems, confining it
around a desirable state, an effort is required to suppress the
intrinsic noise. This noise suppression task entails a cost. For
the important case of thermal noise in an overdamped system, we
show that the minimum cost is achieved when the system control
parameters are held constant: any additional deterministic or
random modulation produces an increase of the cost. We discuss the
implications of this phenomenon for those overdamped systems whose
control parameters are intrinsically noisy, presenting a case
study based on the example of a Brownian particle optically
trapped in an oscillating potential.
\end{abstract}

\pacs{05.40.-a, 02.50.-r}
\keywords{noisy systems, noise suppression, fluctuation phenomena, Brownian motion, It\^{o} calculus}
\maketitle

\section{INTRODUCTION}

Noisy systems are ubiquitous in natural and engineered phenomena.
The presence of noise becomes particularly evident when we move
down into molecular-scale phenomena: the thermal noise,
responsible for the Brownian diffusion of particles, is
omnipresent. However, noise is also intrinsic to many macroscopic
systemsÊ\cite{Sornette2003,Benton2005,Reichenbach2007,Vainstein2007}:
stock markets, population dynamics, and traffic flows, all present
some degree of noise.

Even though in the last years the constructive role of noise has
been appreciated in many physical, chemical, and biological
phenomena -- examples include stochastic resonance
\cite{Benzi1981,Fauve1983,McNamara1988,Wiesenfeld1995,Gammaitoni1998},
noise induced transitions
\cite{NoiseInducedTransitions,Reimann2002}, noise induced
transport \cite{Jlicher1997,MacDonald2000}, stochastic resonant
damping \cite{Volpe2008} --, there are many situations in which
the intrinsic noise of a system is still a nuisance that one wants
to keep under control, and minimize if possible \cite{Vilar2001}.

Noise suppression is a crucial task at all scales.
Microscopic and nanoscopic phenomena have to deal with thermal noise.
Complex pricing systems, such as the Black-Scholes option pricing model \cite{Black1973},
have been developed for dealing with the noise present in the stock markets.
Given the insolubility of the multi-body problem, noise has to be dealt with in the planning of satellites' trajectories.
In all these cases, one needs to exert some kind of
control on the system in order to minimize its intrinsic noise.
Often these actions are controlled by some \textit{input parameters},
which may also vary over time either deterministically or randomly.

Here we will focus on an Ornestein-Uhlenbeck equation, which
describes an overdamped system. Such equation describes a very wide
class of systems, and it has successfully been applied to systems
as diverse as macromolecules that follow the Hooke's law, Brownian
particles, electronic devices, and mesoscopic chemical reactions
\cite{Stochastic}. As a simple example let us consider the
diffusion of a Brownian particle. The particle position variance
is reduced if the particle is confined in a potential well. This
potential well can be produced by various means: by a molecule
that binds the particle, by hydrodynamic focusing, or by optical
or magnetic tweezers. All these means have in common that they
exert a restorting force on the particle whenever it is displaced
from the desired position. A tighter confinement of the particle
is achieved by increasing the stiffness of the link, but a higher
stiffness implicates undergoing an higher cost to run the system.

Here we analyze the cost of noise suppression in non-equilibrium
systems and introduce a \textit{cost function} to quantify the
effort done to control such a system. We show that the minimum
cost is achieved when the system control parameters are held
constant. We find that any additional deterministic or random
modulation of the control parameters entails an increase of the
cost function.

\section{Model}

We consider a dynamical system driven by a Gaussian white random process $dB_t$ (Wiener process),
whose state $s_t$ freely evolves according to $ds_t = dB_t$.
By introducing a restoring force, characterized by a constant stiffness $\bar{k}$, the system can be
forced to fluctuate around a state $\bar{a}$, with variance $\sigma_s^2 = 1/2\bar{k}$.
The stochastic dynamics of the system (known as Ornstein-Uhlenbeck process) is described by
\begin{equation}
ds_t = -\bar{k} (s_t-\bar{a}) dt + dB_t.
\end{equation}

We can now analyze the effect of fluctuating parameters by letting
the mean state $a_t$ and the stiffness $k_t$ -- and therefore the
intrinsic fluctuations $\sigma_s^2 = 1/2k_t$ -- as generic
processes independent of $dB_t$ and possibly dependent between
themselves with $E[a_t] = 0$ and $E[k_t] > 0$, where $E[\cdot]$
denotes expected value. These two conditions guarantee the long
term stability of the system. In the case of a Brownian particle,
they assure that the particle will not eventually escape from the
potential. The former condition ($E[a_t] = 0$), in particular,
signifies that the potential keeps on oscillating around the state
$s = 0$; the latter ($E[k_t] > 0$) that the average stiffness is
positive. In particular, all our conclusions apply to the case in
which $a_t$ and $k_t$ are functions of an Ornstein-Uhlenbeck
process. Notice also that the conclusions also apply to the case
in which $a_t$ and $k_t$ are deterministic function, considering
that these are a special case of random function. The system
time-evolution obeys the equation:

\begin{equation}
ds_t = -k_t (s_t - a_t ) dt + dB_t.\label{e:system}
\end{equation}

As we have already mentioned, the diffusion of a Brownian particle
\textbf{in a time-varying potential} is an example of processes described by (\ref{e:system})
\cite{Volpe2008,Seol2004,Lesanovsky2007}. A free particle diffuses
in such a way that the variance of its position grows linearly
with time. The diffusion process can be partially suppressed by
confining the particle in a potential well which, for example, can
be produced by an optical trap \cite{Neuman2004}. In the presence
of an optical trap, whose center and stiffness oscillate, the
particle position obeys to equation (\ref{e:system}), with $a_t$
being the center of the trap and $k_t$ its stiffness.

With our analysis we aim to find the output variance of the state
of the system described by (\ref{e:system}) when the parameters
vary over time in an arbitrary fashion. In particular we will
identify four contributions in the total variance: \emph{intrinsic
variance} $\sigma^2_{s(i)}$; \emph{stiffness variance}
$\sigma^2_{s(k)}$; \emph{equilibrium variance} $\sigma^2_{s(a)}$;
\emph{interplay covariance} $\rho_{s(ak)}$. All these cases may be
studied experimentally.

Instead of attacking the general case directly, we will proceed by
steps, starting by investigating some limiting cases. This will
permit us to build up the necessary intuition and to gain useful
insights into the phenomenon.

We note that in the cases we study the It\^{o} and Stratonovich
approaches to stochastic integration are mathematically
equivalent, because the diffusion term is constant \cite[pp.
35-37]{Oksendal}. Here we are considering the system's steady
state, but the conclusions apply with little variations also to
the transient.

\subsection{Stationary case ($a_t  \equiv 0$ and $k_t  \equiv \bar{k}$)}

The simplest case is when the equilibrium position of the harmonic
potential does not oscillate ($a_t  \equiv 0$) and its stiffness
is kept constant ($k_t  \equiv \bar{k} >0$). This is the benchmark
against which all other results will be compared. Equation (2)
simplifies as
\begin{equation}
ds_t = -\bar{k} s_t dt + dB_t.
\end{equation}
Its solution can be find multiplying by the integrating factor
$e^{\bar{k}t}$ and comparing with $d\left(e^{\bar{k}t} s_t\right)
= \bar{k} e^{\bar{k}t} s_t dt + e^{\bar{k}t} ds_t$. The solution
is
\begin{equation}
s_t = \underbrace{e^{-\bar{k} t} x_0}_{\rightarrow 0} + e^{-\bar{k} t} \int_0^t e^{\bar{k} u} dB_u
\rightarrow \int_0^t e^{-\bar{k} (t-u)} dB_u,
\end{equation}
where the limit has been taken for large $t$.

It follows that the mean of the system is
$E\left[s_t\right] = 0$
because it is an It\^{o} integral,
and its variance is
\begin{equation}
E[s_t^2] = \int_0^t E\left[e^{-2\bar{k} (t-u)}\right] ds = \int_0^t e^{-2\bar{k} (t-u)} du =
\frac{1-e^{-2\bar{k} t}}{2\bar{k}} \rightarrow \frac{1}{2\bar{k}},
\end{equation}
where the It\^{o} isometry $E\left[\left(\int_0^t f(u,\omega) dB_u\right)^2\right] =
E\left[\int_0^t f^2(u,\omega) du\right]$ has been used \cite{Oksendal,Durrett}.

We can therefore identify the \emph{intrinsic variance} as a
contribution to the total variance of the system:
\begin{equation}
\sigma^2_{s(i)} = \frac{1}{2\bar{k}}.
\end{equation}
As we will see this is the minimum variance that can be achieved
for a given value of the cost function, \emph{i.e.} for a given
$\bar{k}$.

\subsection{Oscillating $k_t$ ($a_t  \equiv 0$)}

When $a_t  \equiv 0$, equation (2) simplifies as
\begin{equation}
ds_t = -k_t s_t dt + dB_t,
\end{equation}
where $k_t$ is an It\^{o} process independent of $B_t$.
Again the solution can be calculated multiplying by the integrating factor $e^{\int_0^t k_u du}$ and comparing with
$d\left(e^{\int_0^t k_u du} s_t\right) = k_t e^{\int_0^t k_u du} s_t dt + e^{\int_0^t k_u du} ds_t$.
Its solution is
\begin{equation}
s_t = \underbrace{e^{-\int_0^t k_u du} x_0}_{\rightarrow 0} + e^{-\int_0^t k_u du} \int_0^t e^{\int_0^u k_v dv} dB_u
\rightarrow \int_0^t e^{-\int_u^t k_v dv} dB_u,
\end{equation}
where the first term vanishes for large $t$ because $E[k_t] = \bar{k}>0$.

We can therefore calculate the mean and the variance of the system.
\begin{equation}
E\left[s_t\right] = E\left[ \int_0^t e^{-\int_u^t k_v dv} dB_u \right] = 0,
\end{equation}
because it is an It\^{o} integral,
and
\begin{equation}
E[s_t^2] = \int_0^t E\left[e^{-2\int_u^t k_v dv}\right] du \ge \int_0^t e^{-2\bar{k} (t-u)} du = \sigma^2_{s(i)}.
\end{equation}
where we have used the It\^{o} isometry and Jensen inequality
\cite{Williams} $E\left[e^{-2\int_u^t k_v dv} du\right] \ge e^{-2
E[k_t] (t-u)}du$ integrated over time with $E[k_t] = \bar{k}$.

We can now identify \emph{the stiffness variance} as a
contribution to the total system variance, caused by the variation
of the stiffness:
\begin{equation}
\sigma^2_{s(k)} = \int_0^t e^{-2\bar{k} (t-u)} E\left[e^{-2\int_u^t (k_v-\bar{k}) dv} - 1 \right] du.
\end{equation}

\subsection{Oscillating $a_t$ ($k_t  \equiv \bar{k}$)}

The case when the equilibrium position of the potential $a_t$ is
oscillating, while $k_t \equiv \bar{k}$ remains constant, was
investigated both theoretically and experimentally in
\cite{Volpe2008}. However, there a different approach was applied
and it can be useful to obtain the same result expressed in the
current formalism. Equation (2) becomes:
\begin{equation}
ds_t = -\bar{k} (s_t-a_t) dt + dB_t.
\end{equation}
It can again be solved by multiplying by the integrating factor integrating factor
$e^{\bar{k}t}$ and comparing with $d\left(e^{\bar{k}t} s_t\right) = \bar{k} e^{\bar{k}t} s_t dt + e^{\bar{k}t} ds_t$.
Its solution is
\begin{equation}
s_t =
\underbrace{e^{-\bar{k}t} s_0}_{\rightarrow 0}
+ \bar{k} e^{-\bar{k}t} \int_0^t e^{\bar{k} u} a_u du
+ e^{-\bar{k}t} \int_0^t e^{\bar{k} u} dB_u
\longrightarrow
\bar{k} \int_0^t e^{\bar{k} (t-u)}  a_u du
+ \int_0^t e^{\bar{k} (t-u)} dB_u.
\end{equation}

Since the process $a_t$ is independent of $B_s$, in the calculation of the
variance of $s_t$ the contributions of the two integrals can be separated
\begin{equation}
E[s_t^2] =  \sigma^2_{s(i)} + \sigma^2_{s(a)},
\end{equation}
where the equilibrium variance
\begin{equation}
 \sigma^2_{s(a)} = \bar{k}^2 E\left[ \left( \int_0^t e^{\bar{k} (t-u)} a_u du \right)^2 \right]
\end{equation}
is the contribution to the variance of the system due to the
oscillation of the equilibrium position of the potential. The
second term is the one corresponding to the stationary state. More
details and a discussion of how this effect produces the
stochastic resonant damping can be found in \cite{Volpe2008}.

\subsection{General case - oscillating $a_t$  and $k_t$}

In the general case given by equation (2), again the solution can be calculated
multiplying by the integrating factor $e^{\int_0^t k_u du}$ and comparing with
$d\left(e^{\int_0^t k_u du} s_t\right) = k_t e^{\int_0^t k_u du} s_t dt + e^{\int_0^t k_u du} ds_t$.
The general solution is
\begin{equation}
s_t =
\underbrace{e^{-\int_0^t k_u du} s_0}_{\rightarrow 0}
+ e^{-\int_0^t k_u du} \int_0^t e^{\int_0^u k_v dv} k_u a_u du
+ e^{-\int_0^t k_u du} \int_0^t e^{\int_0^u k_v dv} dB_u
\longrightarrow
\int_0^t e^{\int_u^t k_v dv} k_u a_u du
+ \int_0^t e^{\int_u^t k_v dv} dB_u.
\end{equation}

For large $t$ following a procedure similar to the previous cases
and finally the variance of the system in the general case is
given by
\begin{eqnarray}\label{total}
E[s_t^2]
&=& E\left[ \left( \int_0^t e^{-\int_u^t k_v dv} dB_u \right)^2 \right] +
E\left[ \left( \int_0^t e^{\int_u^t k_v dv} k_u a_u du \right)^2 \right]\\
&=& \sigma^2_{s(i)} + \sigma^2_{s(k)} +
E\left[ \left( \int_0^t e^{\int_u^t k_v dv} k_u a_u du \right)^2 \right]\\
\label{eqn:variance}
&=&  \sigma^2_{s(i)} + \sigma^2_{s(k)} + \rho_{s(ak)} + \sigma^2_{s(a)},
\end{eqnarray}
where
\begin{equation}
\rho_{s(ak)} = E\left[ \left( \int_0^t e^{\int_u^t k_v dv} k_u a_u du \right)^2 \right] -
\bar{k}^2 E\left[ \left( \int_0^t e^{\bar{k} (t-u)} a_u du \right)^2 \right]
\end{equation}
is the \emph{interplay covariance}, which can be either positive
or negative. However, the total variance is always larger than the
intrinsic variance, since, as it can be seen from equation
(\ref{eqn:variance}), the overall contribution due to the
oscillation of the stiffness and the equilibrium position is
always positive:
\begin{equation}
\sigma^2_{s(k)} + \rho_{s(ak)} + \sigma^2_{s(a)} > 0.
\end{equation}

\subsection{Cost function}

As Eq. (\ref{total}) shows that one can use different protocols in
order to change the output variance of a given intrinsically noisy
system by means of the modulation of the system parameters. Now we
do the next step and we ask the key question of this study: how
one can compare the protocols from the point of view an effort
applied to change the output variance? To deal with such a
question mathematically, we suggest to introduce a \emph{cost
function}.

The idea of a cost function, sometimes referred to as
\emph{objective function}, is very well established in the fields
of economic optimization \cite{OptimizationFinance} and in
engineering \cite{OptimizationEngineering}: it permits one to
compare the performance of systems that work under different
conditions. Typically for a given cost one looks for the
parameters that provide the best performance (the smallest
variance in our case). We introduce here the idea and the
importance of a cost function in the study the confinement of
overdamped systems. An appropriate cost function needs to describe
the overall effort spent in a system to achieve its confinement.

For a stationary system, the stiffness $\bar{k}$ fully describes
the confinement effort. Indeed, as we have seen, the output
variance of a stationary system $\sigma^2_{s(i)}$ is inversely
proportional to the stiffness. Therefore to define the cost
function as $\mathcal{C} = \bar{k}$ seems rather natural.

We introduce a similar cost function for systems whose parameters
are modulated over time. First, as it was shown in
\cite{Volpe2008}, the modulation of the mean state $a_t$ does not
affect the effort made to confine the system; we therefore need to
consider only the modulation of the stiffness in order to
introduce the cost function. For the systems where the stiffness
$k_t$ varies over time, we suggest to use as a cost function the
average value of the stiffness :
\begin{equation}
\bar{\mathcal{C}} = E[k_t].
\end{equation}
 As seen the cost function of a stationary system calculated using
 this formula has the same value as it was defined before.

Let us compare the variance of a stationary system and the same
system but with modulated parameters, assuming that the cost
functions are equal for both systems.  To maintain the cost
function of the system to be constant, we must keep the average
stiffness invariant $E[k_t]$=$\bar{k}$. From this condition it
follows that the intrinsic variance $\sigma^2_{s(i)}$ is also
constant, and, as a straightforward consequence of equation
(\ref{eqn:variance}), it coincides to the minimum variance. This
means that for a given value of cost function the output variance
of the system with modulated parameters is bounded by its
intrinsic variance:
\begin{equation}
E[s_t^2] = \sigma^2_{s(i)} + \sigma^2_{s(k)} + \rho_{s(ak)} +
\sigma^2_{s(a)} > \sigma^2_{s(i)}.
\end{equation}
This can equivalently be stated as the fact that, for a given cost
function, any additional deterministic or random modulation
implicates a larger system variance. For a given value of the cost
function the minimum of the variance is achieved when the control
parameters are constant. From another point of view it means also
that for a given system variance, any additional deterministic or
random modulation produces an increase of the cost function.

\section{Brownian particle example}

Experimental results related with our study were presented in
\cite{Seol2004} where the dynamics of a Brownian particle held in
an optical trap with modulated position and stiffness was
measured. For a given experimental configuration the stiffness of
the (stationary) trap, and, therefore, the achieved confinement is
proportional to the optical power $P$ used to create the trap
($\bar{k} \propto P$). Therefore a higher confinement requires an
higher optical power and the cost function of the system is
defined as $\mathcal{C} = \bar{k} \propto P$.

In the presence of an optical trap, whose center and stiffness
oscillate, the particle position obeys to equation
(\ref{e:system}), with $a_t$ being the center of the trap and $k_t
= 1/\sigma_s^2$ its stiffness. In this case, the cost function is
$\tilde{\mathcal{C}} = E[k_t] \propto E[P(k_t)]$, where the
optical power $P(k_t)$ that must be used to create the optical
trap also fluctuates.

By using of a specific protocol of modulation of the trap
parameters a reduced variance of the observed particle position as
compared to a stationary trap was observed. We analyzed the
experimental data calculating the cost function for the stationary
and modulated traps. When a stationary trap was used (stiffness
$k^{(0)} =3.7\, pN/\mu m$) the output variance of the particle
position was $\sigma_{out}^2 = 1087 \, nm^2$ (Fig. 1(b) of
\cite{Seol2004}). With the oscillating trap, the output position
variance was indeed reduced to $\sigma_{out}^2 = 764 \, nm^2$
(Fig. 1(c) of Ref. \cite{Seol2004}). From the data presented in
\cite{Seol2004} we calculated the average stiffness as $\bar{k} =
E[k_t] = \int_{-\infty}^{+\infty} k p_k(k) dk$, therefore
substituting the time-average with the average over $p_k(k)$ the
probability density of $k_t$ (Fig. 1(b) of Ref. \cite{Seol2004}),
which can be computed from equations (2) and (3) of Ref.
\cite{Seol2004}. The average stiffness of the modulated trap (and
therefore the average optical power introduced in to the system
and the cost function of the process of the noise reduction) was
found considerably bigger that in the stationary case ($\bar{k} =
6.8\, pN/\mu m$). Therefore, we can conclude that the higher
confinement of the particle position in the trap was achieved not
due to the addition of noise to the trapping parameters, but to
the higher average trapping power, while the added noise slightly
increases the output variance. We notice that a stationary trap
with such average power and the same experimental configuration
would produce even smaller output variance than the modulated trap
did.

\section{Conclusions}

We have shown that to suppress the intrinsic noise of a system
entails a cost. For a noisy system in the overdamped regime
controlled by a fluctuating input parameter, we have found out
that the minimum cost is achieved when the system control
parameters are held constant: any additional deterministic or
random modulation of the parameters yields an increase of the cost
function. It is not possible to reduce the output noise of a
system below a threshold value, corresponding to constant input
parameters without increasing the cost function of the system.
This has important implications both from the fundamental point of
view, in order to understand many natural phenomena - for example,
the natural optimization of cellular molecular phenomena -, and
from the engineering one, where it can give a guidance in the
management of the intrinsic noise of a system.

\begin{acknowledgments}
We acknowledge L. Torner for his support to this study. This
research was carried out with the financial support of the Spanish
Ministry of Education and Science (NAN2004-09348-C04-02) and of
the DGCYT under Grant No. FIS2005-01299. It was also partially
supported by the Departament d'Universitats, Recerca i Societat de
la Informaci\'{o} and the European Social Fund. JW thanks Lluis
Torner and Maciej Lewenstein for their hospitality at ICFO.  His
work was partially funded by the NSF grant DMS 0623941.
\end{acknowledgments}


\end{document}